\documentclass[a4paper,twoside]{article}

\usepackage{epsfig}
\usepackage{subcaption}
\usepackage{calc}
\usepackage{amssymb}
\usepackage{amstext}
\usepackage{amsmath}
\usepackage{amsthm}
\usepackage{multicol}
\usepackage{pslatex}
\usepackage{apalike}
\usepackage{SCITEPRESS}     
\usepackage[utf8]{inputenc}

\begin{document}

\title{Identifying the $k$ Best Targets for an Advertisement Campaign via Online Social Networks}

\author{\authorname{Mariella Bonomo\sup{1}, Armando La Placa\sup{1} and Simona E. Rombo\sup{1}}
\affiliation{\sup{1}Department of Mathematics and Computer Science, University of Palermo, Palermo, Italy}
\email{\{mariella.bonomo, armando.laplaca\}@community.unipa.it, simona.rombo@unipa.it}
}


\abstract{We propose a novel approach for the recommendation of possible
customers (users) to advertisers (e.g., brands) based on two main aspects: {\em (i)} the comparison between On-line Social Network profiles, and {\em (ii)} neighborhood analysis on the On-line Social Network. Profile matching between users and brands is considered based on bag-of-words representation of textual contents coming from the social media, and measures such as the Term Frequency-Inverse Document Frequency are used in order to characterize the importance of words in the comparison. The approach has been implemented relying on Big Data Technologies, allowing this way the efficient analysis of very large Online Social Networks. Results on real datasets show that the combination of profile matching and neighborhood analysis  is successful in identifying the most suitable set of users to be used as target for a given advertisement campaign.}
\onecolumn \maketitle \normalsize \setcounter{footnote}{0} \vfill

\section{\uppercase{Introduction}}
\label{sec:introduction}

Social media have gained growing popularity in the last few years, especially On-line Social Networks (OSNs) that enable people to introduce themselves, discuss about their preferred topics, establish and maintain
social connections. This leads advertisers to invest more effort into communicating with
consumers through on-line social networking, that provides suitable platforms for advertising and
marketing. An important issue in this context is how to optimize the effects of marketing communication, by taking advantage of the opportunities offered by the OSNs. 

In particular, advertisers aim to involve in their campaigns those potential consumers who are the most likely interested ones and, hopefully, could spread the received advertisements to other interested users. Automatic systems able to suggest a set of target users for advertising campaigns provide three main benefits: (i) minimization of costs for the dissemination of the advertising campaign through social media, which is often very expensive; (ii) improvement of the user experience in OSNs, since only the possibly interested customers are contacted with advertisements which could be useful for them; (iii) avoid the spread of unuseful information through OSNs.

Here we propose a novel approach for the recommendation of the $k$ best possible consumers to be suggested as target for a specific advertisement campaign. The recommendation is based, on one hand, on the comparison between the OSN profiles associated to users (possible customers) and advertisers (e.g., brands), according to the considered campaign. On the other hand, also the chance that a specific user may distribute the received advertisement to other interested users is considered. In particular, bag of words are used to represent user profiles and profile matching is applied relying on the Term Frequency-Inverse Document Frequency (TF-IDF), in order to weight the importance of the words inside the text associated to user profiles. Moreover, for all users of the considered OSN, their {\em neighbors} and corresponding profiles are taken into account as well, in order to understand to which extent it is convenient sending the advertising to them. 

We have applied our approach to real datasets, which construction is part of the contributions presented here. Indeed, OSN datasets exist which are publicly available but they usually include only network topology, without extensive information on user interests and other related information. On the other hand, complementing the available network topologies via web-scraping starting from personal access points is not trivial and often not possible. We present a methodology for the association of contents to the nodes of a OSN, given its topology, based on following the cross-linked references of public web-pages. This allows to fulfil that neighbor nodes in the network may share common contents more likely than nodes very far each other. The obtained results are promising, indeed the approach allows to correctly identify the most suitable users in all the considered situations.

\section{\uppercase{Related Work}}

\noindent 
Modeling the user profiles from social media raw data is usually a challenging task. The approaches proposed in the Literature to this aim may be roughly classified in two main categories. The first category includes approaches based on the analysis of user generated contents (here referred to as {\it semantic approaches}). As for the approaches in the second category, individuals are characterized by ''actions'', e.g., visited web pages ({\it action-based approaches}). Our framework belongs to the first category.

\paragraph{Semantic approaches}

The authors of \cite{Schwartz2013} use Differential Language Analysis (DLA) in order to find language features across millions of Facebook messages that distinguish demographic and psychological attributes. They show that their approach can yield additional insights (correlations between personality and behavior as manifest through language) and more information (as measured through predictive accuracy) than traditional a priori word-category approaches. 

The framework proposed in \cite{Lin2014} relies on a semi-supervised topic model to construct a representation of an app's version as a set of latent topics from version metadata and textual descriptions. The authors discriminate the topics based on genre information and weight them on a per-user basis, in order to generate a version-sensitive ranked list of apps for a target user.

In \cite{LiangZRK18} the authors propose a dynamic user and word embedding algorithm that
can jointly and dynamically model user and word representations in the same semantic space. They consider the context of streams of documents
in Twitter, and propose a scalable black-box variational inference algorithm
to infer the dynamic embeddings of both users and words in streams. They also propose a streaming keyword diversification model to diversify top-K keywords for characterizing users’ profiles over time.

The first technique applied to brand-affinity matching that is not an action-based approach has been presented in \cite{BonomoCSR19}. In particular, the authors present a profile-matching technique based on tree-representation of user profiles and apply it on Facebook ego-networks. The approach presented here extends those results, showing that a suitable combination of profile-matching and neighborhood analysis is more successful in identifying the best $k$ users for advertisements distribution.

\paragraph{Action-based approaches}
In \cite{Provost2009} individuals are associated each other due to some actions they share (e.g., they have visited the same web pages). The proximity between individuals on networks built upon such relationships is informative about their profile matching. In particular, brand-affinity audiences are built by selecting the social-network neighbors of existing brand actors, identified via co-visitation of social-networking pages. This is achieved without saving any information about the identities of the browsers or content of the social-network pages, thus allowing for user anonymization.

In \cite{Ahmed2011} compact and effective user profiles are generated from the history of user actions, i.e., a mixture of user interests over a period of time. The authors propose a streaming, distributed inference algorithm which is able to handle tens of millions of users. They show that their model contributes towards improved behavioral targeting of display advertising relative to baseline models that do not incorporate topical and/or temporal dependencies. 

 In \cite{Iglesias2011} a computer user behavior is represented as the sequence of the commands she/he types during her/his work. This sequence is transformed into a distribution of relevant subsequences of commands in order to find out a profile that defines its behavior. Also, because a user profile is not necessarily fixed but rather it evolves/changes, the authors propose an evolving method to keep up to date the created profiles using an Evolving Systems approach.

The observation that behavior of users is highly influenced by the behavior of their neighbors or community members is used in  \cite{XieLMLCR14} to enrich user profiles, based on latent user communities in collaborative tagging.


\section{\uppercase{Proposed Approach}}
\noindent
The main goal of the proposed approach is to identify the most suitable $k$ possible {\it buyers} to whom distributing a given advertisement campaign. To this aim, two important aspects have to be taken into account:

\begin{itemize}
    \item Ideally, users to whom distributing the campaign should have interests compatible with the specific features of the advertiser (i.e., the {\it brand}).
    \item It would be better if the chosen possible buyers would know other users whose interests are close to those expected for the campaign success as well.
\end{itemize}

The second point has a twofold effect. Indeed, it can be useful in order to obtain the {\it next} $k$ users to contact and, at the same time, maximize the chance that they can be contacted directly by the buyers selected at the current step.

In order to accomplish the two above points, we have based the presented research on the use of Online Social Networks. 

\paragraph{Online Social Network (OSN)}
We represent an Online Social Network as an undirect graph ${\cal N}=(V,E)$ such that nodes in $V$ are associated to the users and two nodes are linked in ${\cal N}$ if a social relationship (e.g., friendship, common interests, etc.) between them occurs in the represented OSN. In addition to the topological representation of an OSN, further details are necessary in order to characterize each node. 

\paragraph{User Profile}
User profiles complement network topology information. In particular, each node in the network points to data associated to a user and retrieved from the considered social media. What is important for this research, is textual information about user general interests and activities, coming for example from private communications, posts, comments, short text messages. Therefore, the user profile of $u$ is represented here by a text $T_u$ characterizing $u$ with references to the considered OSN.

\paragraph{Brand profile} Also a brand profile is represented by a text, that can be for example easily extracted from the web-page describing brand activities or from other textual documents containing information on the advertisement campaign. In the following, we refer indistinctly to brand profile and advertisement campaign, since both may be described by textual documents and then handled in the same way in the context of the proposed approach.

\subsection{Profile Matching}
Let $u$ be a node in an input OSN ${\cal N}$ and $T_u$ be its user profile. Moreover, let $T_b$ the text associated to the brand profile. The first step of our approach is to understand how much $T_u$ and $T_b$ are "similar", i.e., to which extent they {\em match} each other. To this aim, we consider TF-IDF and cosine similarity measures in order to understand if and how much textual contents associated to $u$ and to the brand are semantically related. This is sketched in the following, for the specific case under consideration (i.e., only two textual documents, $T_u$ and $T_b$).

\paragraph{Importance of words} Let $w_{ij}$ be a word occurring in the text $T_j$ ($j=1, \ldots, m$). The TF-IDF function for $w_{ij}$ is defined as: $$TF\text{-}IDF(w_{ij})=TF(w_{ij})*IDF(w_{ij})$$

\noindent such that: 

$$TF(w_{ij})=\frac{|w_{ij}|}{|T_j|}$$ 

\noindent where $|w_{ij}|$ is the length of $w_{ij}$ and $|T_j|$ is the number of words in $T_j$, and: $IDF(w_{ij})=\log \frac{m}{h}$, where $h \leq m$ is the number of texts where $w_{ij}$ occurs.

\paragraph{Affinity between profiles} Let $V_u$ and $V_b$ be two arrays of $k$ real values. The cosine similarity between $V_u$ and $V_b$ is defined as: $$\mathcal{CS}(V_u,V_b)=\frac{\sum_{i=1}^{k}V_u[i]*V_b[i]}{\sqrt{\sum_{i=1}^k V_u[i]^2} * \sqrt{\sum_{i=1}^k V_b[i]^2}}$$

\noindent The {\em affinity} between the profiles associated to an user and a brand is then computed as the cosine similarity between arrays containing the TF-IDF values of the words occurring in $T_u$ and $T_b$:

$$\mathcal{A}(T_u,T_b)=\mathcal{CS}(V_u,V_b)$$

\subsection{Neighborhood Analysis}
In order to make more effective the advertisement campaign, for each node $u$ in $V$, it is important not only to measure to what extent its profile matches with the brand profile, but also how many nodes in the neighborhood of $u$ could be possibly interested in that campaign as well. That is, the best targets are those nodes which profile matches with the brand, and that are surrounded by other nodes with this same feature.

\paragraph{Node centrality} Let $u$ be a node in the set of vertices $V$ and $T_u$ and $T_b$ be the user and brand profiles, respectively. Moreover, let $N_u$ be the set of nodes linked to $u$ by at least one edge in the set of edges $E$ of $\cal{N}$. Then, the {\em centrality} of $u$ for the given considered brand (or advertisement campaign) is defined as:

$$\mathcal{C}(u,b) = \frac{{\sum}_{v\in N_u} \mathcal{A}(T_v,T_b)}{|N_u|}$$

\noindent It is worth pointing out that, in order to focus the advertising campaign on those interested users only, a threshold value can be chosen on the affinity values according to which filtering only nodes in the network scoring affinity values larger than that threshold. 

\paragraph{Node utility} As already explained, the final aim of our approach is to identify the best $k$ nodes to which distribute advertisements according to their profile matching with the brand (or campaign, respectively). On the other hand, in order to maximize the gain, we are also interested into detect nodes which neighbors in the OSN may be interested in the same advertisements. To this respect, the {\em utility} of a node for a specific brand/campaign is defined as follows:

$$\mathcal{U}(u,b) = \alpha \cdot \mathcal{A}(T_u,T_b) + (1-\alpha)\cdot \mathcal{C}(u,b)$$

\noindent where $\alpha$ is a real value in $[0,1]$ used to balance two different contributions, i.e., the match between user and brand, and the match between user neighbors and brand.

\subsubsection{Example}
\begin{figure}[ht]
\centering
\includegraphics[width=7cm]{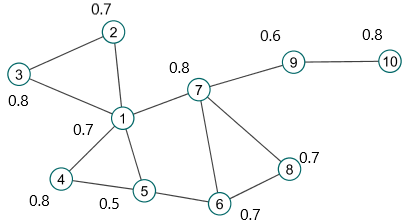}
\caption{A small OSN. For each node, the corresponding affinity value is also shown.}
\label{fig::example}
\end{figure}

Figure \ref{fig::example} depicts a small OSN and, for each node, the corresponding affinity value that has supposed to be computed with respect to a given brand is also shown. Suppose that the brand is interested to send its advertising campaign to $5$ nodes on this network (i.e., $k=5$).

\noindent As an example, for Node $1$:

$$ \mathcal{C}(1,b)={\frac{0.7+0.8+0.8+0.5+0.8}
	{ 5 }}=0.72$$

\noindent and, for $\alpha= 0.4$:

$$ \mathcal{U}(1,b)=0.4 \cdot 0.7 + (1-0.4)\cdot 0.72=0.71$$

\noindent while for $\alpha= 0.6$:

$$ \mathcal{U}(1,b)=0.6 \cdot 0.7 + (1-0.6)\cdot 0.72=0.7$$

\noindent Tables \ref{tab::04}-\ref{tab::06} show the utility and centrality values for nodes in the example OSN, for three different values of $\alpha$. In particular, nodes are sorted according to the utility, and filtered such that only those with utility values larger than $0.6$ are shown in the tables for each considered $\alpha$.

\begin{table}[h]
\centering
    \begin{tabular}{|c|c|c|}
    \hline
	\textbf{Node }& \textbf{utility}& \textbf{centrality}\\
	\hline
    $3$&  $0.74$& $0.7$\\ \hline
    $2$&  $0.73$& $0.75$\\ \hline
    $7$&  $0.72$& $0.67$\\ \hline
    $9$&  $0.72$& $0.8$\\ \hline
    $1$&  $0.71$& $0.72$\\ \hline
    \end{tabular}
    \caption{Top nodes for $\alpha =0.4$, sorted according to their utility values.}
    \label{tab::04}
	\end{table}
	
\begin{table}[h]
\centering
    \begin{tabular}{|c|c|c|}
    \hline
	\textbf{Node }& \textbf{utility}& \textbf{centrality}\\
	\hline
    $3$&  $0.75$& $0.7$\\ \hline
    $7$&  $0.73$& $0.67$\\ \hline
    $2$&  $0.72$& $0.75$\\ \hline
    $8$&  $0.72$& $0.75$\\ \hline
    $4$&  $0.7$& $0.6$\\ \hline
    \end{tabular}
    \caption{Top nodes for $\alpha =0.5$, sorted according to their utility values.}
    \label{tab::05}
	\end{table}
	
\begin{table}[h]
\centering
    \begin{tabular}{|c|c|c|}
    \hline
	\textbf{Node }& \textbf{utility}& \textbf{centrality}\\
	\hline
    $3$&  $0.76$& $0.7$\\ \hline
    $7$&  $0.74$& $0.67$\\ \hline
    $2$&  $0.72$& $0.75$\\ \hline
    $4$&  $0.72$& $0.6$\\ \hline
    $8$&  $0.72$& $0.75$\\ \hline
    \end{tabular}
    \caption{Top nodes for $\alpha =0.6$, sorted according to their utility values.}
    \label{tab::06}
\end{table}

\noindent From Tables \ref{tab::04}-\ref{tab::06}, it is evident that Node $3$ is the best target for the input brand for all considered $\alpha$ values, according to the supposed affinity values. This is due to the fact that Node $3$ and its two neighbors $1$ and $2$ all score good affinity values, therefore Node $3$ keeps its top ranking position both when it is given higher importance to its affinity with the brand (larger $\alpha$ value) and when, instead, the focus is on the affinity of its neighbors (smaller  $\alpha$ value). In the transition from smaller to larger $\alpha$ values, the configuration of top ranking change for the further four positions. From $\alpha=0.4$ to $\alpha=0.5$, Nodes $2$ and $7$ exchange their position in the ranking, due to the fact that both them have neighbors with high affinity values but the affinity between Node $7$ and the brand is higher than that of Node $2$. Another effect is that Nodes $1$ and $9$ are replaced by Nodes $4$ and $8$, respectively. Again, all such nodes have neighbors with high affinity values, but the latter nodes have larger affinity values than the former ones. Analogous considerations can be done from $\alpha=0.5$ to $\alpha=0.6$, where the only effect is the inversion between Nodes $4$ and $8$ in the ranking.

\section{\uppercase{RESULTS}}

The main goal of our experimental validation has been to verify on large OSN datasets if the choice of $k$ best targets according to the measures introduced here is effective. To this aim, a first important problem to be solved has been the construction of the input OSN. Indeed, while a number of social network graphs are publicly available, the same is not true for network users profiles. In the following of this section, we first discuss these aspects related to OSN construction, and then present some results we have obtained by applying our approach on datasets coming from the real world. The proposed approach has been implemented in Java $1.8$ under Apache Spark $1.6$. To this respect, the use of Big Data Technologies allow to exploit the software tool also on very large OSNs.

\subsection{Network construction}
OSN graphs are available for example from Standford website ({\tt https://snap.stanford.edu/data/}). We have considered the {\tt twitter-2010} OSN from that repository, having $90,908$ vertices and $443,399$ edges. Unfortunately, the available OSNs consist only on the graph topology, no information about user interests and profiles are publicly available. Web scraping has been used here in order to collect and extract useful contents for user profiles characterization. In particular, we have avoided to associate randomly the information obtained by web scraping to nodes in the considered OSN graph, due to the fact that a random association would have altered  the natural mechanism according to which users in the same neighbors have similar interests. In order to mimic such a mechanism, which is important for our approach (indeed the introduced measures aim at detecting neighbor nodes with similar interests), we have proceeded as follows.

We have first randomly selected $20$ "seed" nodes from the {\tt twitter-2010} OSN and $20$ web-pages focused on different topics (cooking, fashion, cars, etc.). Indeed, with a certain margin of simplification, we have assumed  that a user profile may be obtained by scraping the contents of a web-page on a specific topic. Then, a visit in depth of the OSN has been performed starting from each of the seeds and stopping when the entire network was visited. For each new node to be visited, a new web-page has been visited as well, following the cross-page links on the considered web-pages.

\subsection{Experimental validation}

Our experimental analysis has been devoted to understand to what extent our approach is effective, in order to identify the $k$ most convenient nodes in the input OSN to which distribute the advertisement. As already explained, the main aim here is to optimize two different aspects when identifying the best targets, that is, the fact that interests of considered users are related to the campaign contents, and the fact that they have ``friends'' on the OSN potentially interested to the distributed advertisements. We have considered the web-pages associated to four brands, listed in Table \ref{tab::brands}. 

\begin{table}[h]
 \centering
\begin{tabular}{|l|l|}
  \hline
  \textbf{Brand}& \textbf{Web-page}\\
  \hline
  AlphaRomeo & {\tt www.alfaromeo.it} \\
  \hline
  Amarelli & {\tt www.amarelli.it} \\
  \hline 
  Carpisa & {\tt www.carpisa.it} \\
  \hline
  KikoCosmetic & {\tt www.kikocosmetics.com} \\
  \hline
\end{tabular}
\caption{The considered brands and their associated web-pages.}\label{tab::brands}
\end{table}

We have considered the OSN constructed as described in the previous section and we have computed, for each of the four brands, the different values of affinity and utility (with $\alpha=0,25; 0,5; 0,75$) for all nodes in the network. Then, we have ranked them in descending order, according to each of these measures. We have supposed that the number of target nodes is $k=100$ and we have fixed to $0.6$ the minimum value of affinity between user and brand profiles in order a user to be considered a possible target.

The obtained results have been compared with a random choice of the $k$ nodes to which distribute the advertisement. For $100$ different times, $100$ nodes have been extracted from the set of vertices $V$ and the affinity between their and brand profiles have been computed at each time. The obtained results for the different brands do not present significant differences, therefore we illustrate only those regarding AlphaRomeo in Table \ref{tab::comparison}. In particular, the considered method is specified in the first column of the table, and for the Random generation we have considered the average of obtained results. For each method, the number of nodes presenting an affinity value larger than the chosen threshold when the first $k$ nodes in the corresponding ranking is chosen is shown in the third column. It is interesting to observe that, with respect to the random choice, both Affinity and Utility with a high value of $\alpha$ ($0.75$) improves of one order of magnitude. Indeed, in this two latter cases, all the considered nodes have affinity values above the threshold. This shows that the profile matching at the basis of our approach is effective in the selection of target users for an advertising campaign. However, the second aspect to take into consideration is related to the number of possible further interested users that can be reached by the advertisement, starting from those $k$. To this respect, the last column of Table \ref{tab::comparison} shows how many distinct nodes are in the neighborhoods of the first $k$ ones (according to the ranking obtained for each method). The second column of the table shows the total number of nodes with affinity values larger than the threshold that can be reached starting from the first $k$, for each ranking. It is evident that, again, the worst performance is obtained by the Random method, whereas the best one by Utility with $\alpha=0.5$ in this case. This confirms what expected, that is, neighborhood analysis associated to profile matching is the most promising choice.

\begin{table*}[h]
 \centering
\begin{tabular}{|c|c|c|c|}
  \hline
  \textbf{Method}& \textbf{$\#$ of Target Nodes}& \textbf{Directly Reached}& \textbf{From Neighborhoods}\\
  \hline
  Affinity & $184$ & $\textbf{100}$ & $84$\\
  Utility ($\alpha=0.25$) & $152$ & $64$ & $88$\\
  Utility ($\alpha=0.5$) & $\textbf{192}$ & $99$ & $\textbf{93}$\\
  Utility ($\alpha=0.75$) & $181$ & $\textbf{100}$ & $81$\\
  Random  & $99$ & $13$ & $86$\\
  \hline
\end{tabular}
\caption{Total number of nodes (second column) with affinity values larger than the chosen threshold identified by each method (first column), fraction of target nodes directly reached (third column) or instead detected from the neighborhoods (fourth column).}\label{tab::comparison}
\end{table*}

\begin{figure}[!ht]
\centering
\includegraphics[width=7cm]{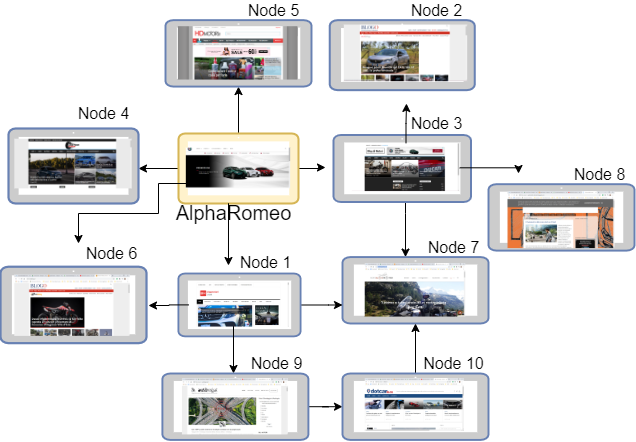}
\caption{Links among the first $10$ target nodes for Alpha Romeo.}\label{tab::alpharomeo_graph}
\end{figure}

\begin{figure}[!ht]

\includegraphics[width=7cm]{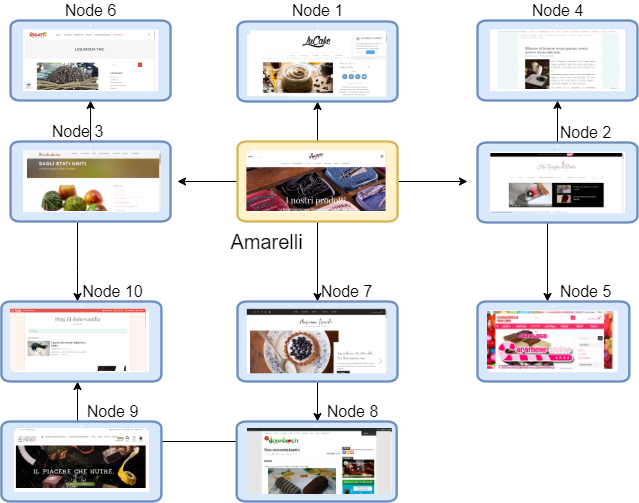}
\caption{Links among the first $10$ target nodes for Amarelli.}\label{tab::amarelli_graph}
\end{figure}

\begin{figure}[!ht]
\centering
\includegraphics[width=7cm]{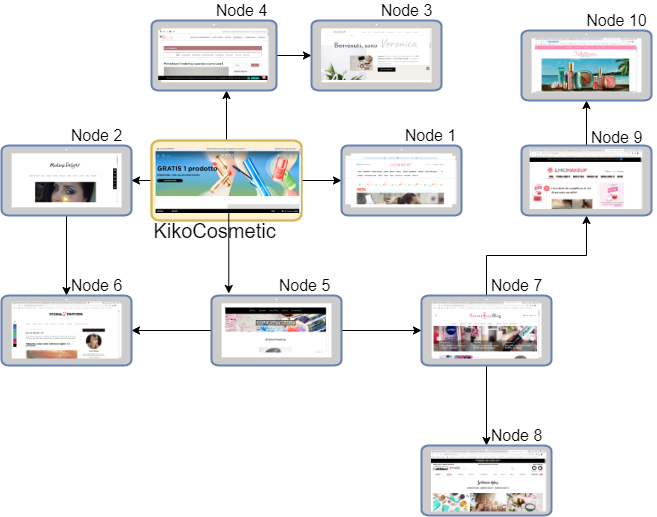}
\caption{Links among the first $10$ target nodes for Kiko Cosmetics.}\label{tab::kiko_graph}
\end{figure}

Figures \ref{tab::alpharomeo_graph}-\ref{tab::kiko_graph}
provide a graphical illustration of the links between the first $10$ target nodes for each of the considered brand according to the method Utility with $\alpha=0.5$. In particular, the web-page of the brand is the central node, and the web-pages associated to the first $10$ nodes in the ranking are depicted around, showing also the existing links among them in the OSN. For all brands, most of the considered target nodes are connected in paths, trees or small communities. 

\begin{figure}[!ht]
\centering
\includegraphics[width=7cm]{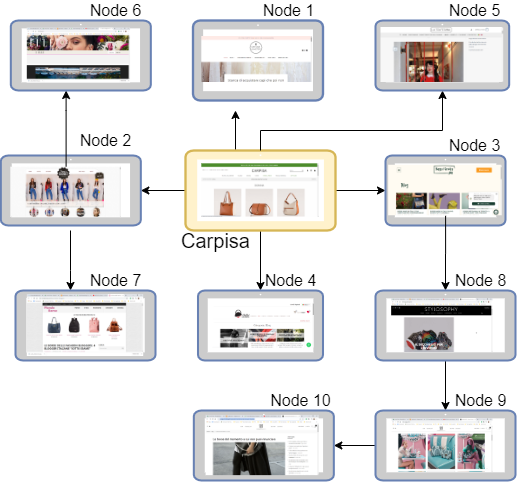}
\caption{Links among the first $10$ target nodes for Carpisa.}\label{tab::carpisa_graph}
\end{figure}

Tables \ref{tab::alpharomeo}-\ref{tab::kiko} show the web-links of the top $10$ nodes for each brand, and their values of utility and affinity. Moreover, in the last column is also reported, for each node, the number of neighbors that are target nodes as well (i.e., their affinity value is larger than the considered threshold). It is evident that the top target nodes refers to web-pages which contents are strictly related with those of the brand, in each case.

\section{\uppercase{Concluding Remarks}}
\label{sec:conclusion}

We have discussed here how the combination of information retrieval measures for profile matching and neighborhood exploration in OSNs may be successful in order to identify a set of target users for the distribution of advertisements. In particular, such users not only have interests related to the contents of the advertisement, but may also potentially spread the received advertisements to other interested users in the OSN. This allows to minimize costs for advertising campaigns, improve user experience in OSNs and avoid spread of unuseful information through OSNs.  

Results obtained by the measures introduced here on real datasets are promising. However, we are conscious that the proposed approach relies on a naive, although effective, technique for neighborhood exploration. In our future work we plan to extend it by taking into account more complex node centrality measures \cite{GIANCARLO2019950,MohammedZSMH20}. Moreover, we will explore the direction of including OSNs community detection \cite{WadhwaB14} in our analysis, in order to identify compact groups of nodes with interests related to the considered input advertising campaigns. 

Finally, we conclude with the following observation. An important problem in the context of OSNs analysis is the absence of publicly available datasets including not only network topology, but also structured information related to the network users, such as interests, general data, actions, etc.. It is worth to point out that the construction of such datasets via web-scraping starting from personal access points on the OSN  presents several problems, among which data privacy constraints, the fact that the obtained networks would be mostly ego-networks \cite{ARNABOLDI201744,KwonMHKMH19}, and the difficulty in building networks that reflect the sizes of real OSNs, often very large \cite{PengWX17,WU202079}.  Therefore, providing suitable OSN public datasets which contain both topological and semantic data would be a valuable contribution for the scientific community. We plan to extend in this direction the procedure described here for the construction of big OSNs, and to provide a public repository containing such datasets.

\section*{\uppercase{Acknowledgements}}
Part of the research presented here has been funded by the MIUR-PRIN  research project ``Multicriteria Data Structures and Algorithms: from compressed  to learned indexes, and beyond'', grant n. 2017WR7SHH, and by the INdAM - GNCS Project 2020 ``Algorithms, Methods and Software Tools for Knowledge Discovery in the Context of Precision Medicine''.

\bibliographystyle{apalike}
{\small
\bibliography{references}}

\newpage

\begin{table*}[t!h]
\centering
\begin{tabular}{|c|c|c|c|c|}
  \hline
  \textbf{Network's node}& \textbf{Brand: Alpha Romeo}& \textbf{Utility}& \textbf{Affinity} &\textbf{$\#$ of Target Nodes}\\
  \hline
  Node 1 & blogmotori.com & 0.84 & 0.64 & 5\\
  \hline
  Node 2 & autoblog.it & 0.80 & 0.61 & 2\\
  \hline
  Node 3 & blogdimotori.it & 0.88 & 0.87 & 5\\
  \hline 
  Node 4 & hdmotori.it/auto & 0.74 & 0.71 & 2\\
  \hline
  Node 5 & motori.news & 0.73 & 0.69 & 2\\
  \hline
  Node 6 & motoblog.it & 0.91 & 0.67 &3\\
  \hline
  Node 7 & motoblogtrotter.com & 0.91 & 0.71 &4\\
  \hline
  Node 8 & automobile-domani.blogspot.com & 0.88 & 0.64 &2\\
  \hline
  Node 9 & autologia.net & 0.80 & 0.62 & 3\\
  \hline
  Node 10 & dotcar.it/ & 0.94 & 0.92 & 3\\
  \hline
\end{tabular}
\caption{First $10$ target nodes for Alpha Romeo.}\label{tab::alpharomeo}
\end{table*}

\begin{table*}[t]
 \centering
\begin{tabular}{|c|c|c|c|c|}
  \hline
  \textbf{Network's node}& \textbf{ Brand: Amarelli}& \textbf{Utility}& \textbf{Affinity}& \textbf{$\#$ of Target Nodes}\\
  \hline
  Node 1 & lucake.it & 0.89 & 0.78 & 2\\
  \hline
  Node 2 & hovogliadidolce.it & 0.93 & 0.86& 4\\
  \hline
  Node 3 & brindando.com & 0.85 & 0.70 & 4\\
  \hline 
  Node 4 & cioccolatoeliquirizia.it & 0.93 & 0.61 & 2\\
  \hline
  Node 5 & caramelleonline.com/blog & 0.75 & 0.79 & 2\\
  \hline
  Node 6 & mentaeliquirizia.com & 0.89 & 0.63 &2\\
  \hline
  Node 7 & veganblog.it & 0.75 & 0.63 & 3\\
  \hline
  Node 8 & cacaocrudo.it/it & 0.87 & 0.69 & 3\\
  \hline
  Node 9 & blog.cookaround.com/dolcevanilia & 0.76 & 0.76 & 3\\
  \hline
  Node 10 & blog.rigato.net/tag/liquirizia & 0.92 & 0.70 & 3\\
  \hline
\end{tabular}
\caption{First $10$ target nodes for Amarelli.}\label{tab::amarelli}
\end{table*}

\begin{table*}[!ht]
\centering
\begin{tabular}{|c|c|c|c|c|}
  \hline
  \textbf{Network's node}& \textbf{Brand: Carpisa}& \textbf{Utility} & \textbf{Affinity}& \textbf{$\#$ of Target Nodes}\\
  \hline
  Node 1 & concosalometto.com & 0.75 & 0.74 & 1\\
  \hline
  Node 2 & ireneccloset.com & 0.71 & 0.61 & 4\\
  \hline
  Node 3 & bagsandfruits.com/it/blog & 0.82 & 0.67 & 2\\
  \hline 
  Node 4 & ilbellodelleborse.com/blog & 0.75 & 0.70 &2\\
  \hline
  Node 5 & latolfetana.com/blog/page/2 & 0.77 &0.66 &2\\
  \hline
  Node 6 & elle.com & 0.79 & 0.61 & 2\\
  \hline
  Node 7 & mondoborse.com & 0.88 & 0.73 & 2\\
  \hline
  Node 8 & bags.stylosophy.it & 0.70 & 0.62 &3\\
  \hline
  Node 9 & saragiunti.it & 0.92 & 0.90 & 3\\
  \hline
  Node 10 & saragiunti.it/blog & 0.75 & 0.68 & 2\\
  \hline
\end{tabular}
\caption{First $10$ target nodes for Carpisa.}\label{tab::carpisa_graph}
\end{table*}

\begin{table*}[!ht]
\centering
\begin{tabular}{|c|c|c|c|c|}
  \hline
  \textbf{Network's node}& \textbf{Brand: KikoCosmetic}& \textbf{Utility}& \textbf{Affinity}& \textbf{$\#$ of Target Nodes}\\
  \hline
  Node 1 & blog.cliomakeup.com & 0.81 & 0.70 & 3\\
  \hline
  Node 2 & makeupdelight.com & 0.62 & 0.61 & 2\\
  \hline
  Node 3 & claudia-makeup.com/blog-trucco & 0.82 & 0.73 & 2\\
  \hline 
  Node 4 & polveredistellemakeup.com & 0.77 & 0.62 & 3\\
  \hline
  Node 5 & aboutbeautymakeup.wordpress.com & 0.85 & 0.68 & 4\\
  \hline
  Node 6 & donnaedintorni.com/blog-makeup & 0.85 & 0.80 & 3\\
  \hline
  Node 7 & loscrigno.it/beautycase & 0.77 & 0.70 & 4\\
  \hline
  Node 8 & sabbioni.it/it/blog.php & 0.85 & 0.71 & 2\\
  \hline
  Node 9 & ilmiomakeup.it & 0.76 & 0.64 & 3\\
  \hline
  Node 10 & follettarosamakeup.com & 0.75 & 0.63 & 2\\
  \hline
\end{tabular}
\caption{First $10$ target nodes for Kiko Cosmetics.}\label{tab::kiko}
\end{table*}




\end{document}